\documentclass[cpc,amsmath,amssymb,showpacs,superscriptaddress,nofootinbib,twocolumn]{revtex4-1}
\usepackage{graphicx}
\usepackage{epstopdf}
\usepackage{dcolumn}
\usepackage{color}
\usepackage{bm}
\usepackage{CJK,CJKnumb,CJKulem}     
\usepackage{overpic}
\usepackage{rotating}
\usepackage{times}
\usepackage{multirow}
\usepackage{float}
\usepackage{mathrsfs}
\usepackage{booktabs}

\newcommand{\pp}{\pi^+\pi^-}

\newcommand{\LL}{\ell^+ \ell^-}
\newcommand{\EE}{e^+e^-}

\newcommand{\psip}{\psi(3686)}

\newcommand{\jpsi}{J/\psi}
\newcommand{\po}{\pi^{0}}
\begin{document}
\hyphenpenalty=10000
\tolerance=1000
\begin{CJK*}{GBK}{song}

\title{\boldmath Measurement of branching fractions for $\psip\rightarrow\gamma\eta',\gamma\eta$ and $\gamma\pi^{0}$}
\author{
  \begin{small}
    \begin{center}
M.~Ablikim$^{1}$, M.~N.~Achasov$^{9,d}$, S. ~Ahmed$^{14}$, M.~Albrecht$^{4}$, M.~Alekseev$^{54A,54C}$, A.~Amoroso$^{54A,54C}$, F.~F.~An$^{1}$, Q.~An$^{51,41}$, J.~Z.~Bai$^{1}$, Y.~Bai$^{40}$
% ...
\\
\vspace{0.2cm}
(BESIII Collaboration)\\
\vspace{0.2cm} {\it
$^{1}$ Institute of High Energy Physics, Beijing 100049, People's Republic of China\\
% ...
$^{4}$ Bochum Ruhr-University, D-44780 Bochum, Germany\\
% ...
$^{9}$ G.I. Budker Institute of Nuclear Physics SB RAS (BINP), Novosibirsk 630090, Russia\\
% ...
$^{14}$ Helmholtz Institute Mainz, Johann-Joachim-Becher-Weg 45, D-55099 Mainz, Germany\\
% ...
$^{40}$ Southeast University, Nanjing 211100, People's Republic of China\\
$^{41}$ State Key Laboratory of Particle Detection and Electronics, Beijing 100049, Hefei 230026, People's Republic of China\\
% ...
$^{51}$ University of Science and Technology of China, Hefei 230026, People's Republic of China\\
% ...
$^{54}$ (A)University of Turin, I-10125, Turin, Italy; (B)University of Eastern Piedmont, I-15121, Alessandria, Italy; (C)INFN, I-10125, Turin, Italy\\
% ...
$^{d}$ Also at the Novosibirsk State University, Novosibirsk, 630090, Russia\\
% ...
}\end{center}
\end{small}
}
\affiliation{}

\date{\today}
\begin{abstract}

Using a data sample of  $448\times10^{6}$ $\psip$ events collected with the BESIII detector operating at the BEPCII storage ring, the decays
$\psip\rightarrow\gamma\eta$ and $\psip\rightarrow\gamma\pi^{0}$ are observed with a statistical significance of $7.3\sigma$ and $ 6.7\sigma$, respectively.
The branching fractions are measured to be $\mathcal{B}(\psip\rightarrow\gamma\eta)=(0.85\pm0.18\pm0.05)\times10^{-6}$ and
$\mathcal{B}(\psip\rightarrow\gamma\pi^{0})=(0.95\pm0.16\pm0.05)\times10^{-6}$.
In addition, we measure the branching fraction of $\psip\rightarrow\gamma\eta'$ to be $\mathcal{B}(\psip\rightarrow\gamma\eta')=(125.1\pm2.2\pm6.2)\times 10^{-6}$,
with improved precision compared to previous results.

\end{abstract}

\pacs{13.20.Gd}
\maketitle
\section{\boldmath Introduction}

Radiative decays to light hadrons comprise a substantial fraction of the decays of vector charmonium states, $e.g.$, 6\% for $\jpsi$ and 1\% for
$\psip$~\cite{Scharre:1980yn} with respect to their total width.
In previous experiments, only about 10\% of the expected $\jpsi$ and $\psip$ radiative decays have been observed exclusively~\cite{PDG}.
Within the framework of Quantum Chromodynamics (QCD), radiative decays of the vector charmonium states proceed predominantly via the emission
of a real photon from the $c$ or $\bar{c}$ quark, followed by the $c\bar{c}$ annihilation into two gluons.

Various phenomenological mechanisms, such as $\eta_{c}$-$\eta^{(\prime)}$ mixing~\cite{mixing1,mixing2},
final-state radiation by light quarks~\cite{VMD1,VMD2}, and the vector-meson dominance model in association with
$\eta_{c}$-$\eta^{(\prime)}$ mixing~\cite{VMD_Mix}, are proposed to explain the properties of charmonium state radiative decays to a pseudoscalar meson.
Measurements of these charmonium radiative decays provide important tests for the different theoretical predictions.
The ratio $R_{\jpsi}\equiv\frac{\mathcal{B}(\jpsi\rightarrow\gamma\eta)}{\mathcal{B}(\jpsi\rightarrow\gamma\eta')}$ has been predicted
based on the first-order perturbative QCD calculation,
and $R_{\psip}\equiv\frac{\mathcal{B}(\psip\rightarrow\gamma\eta)}{\mathcal{B}(\psip\rightarrow\gamma\eta')}$ is expected to be approximately equal to $R_{\psip}\approx R_{\jpsi}$~\cite{R1R2}.
The decay rates of $\jpsi$ and $\psip\rightarrow\gamma\po$ are expected to be smaller than those of $\jpsi$ and $\psip\rightarrow\gamma\eta$ or $\gamma\eta'$ as a consequence of suppressed gluon coupling to isovector currents.
By assuming that the partial widths of $\jpsi\to\gamma\eta$ and $\gamma\eta'$ are saturated by the $\eta_c$-$\eta'$ mixing, the predicted branching fractions of $\jpsi\to\gamma\eta$ and $\gamma\eta'$ were accounted for to the correct orders of magnitude in Ref.~\cite{SATURATE}.

The CLEO experiment~\cite{Br_cleo} measured the branching fractions of $\jpsi$ and $\psip$ decays to $\gamma\po,\gamma\eta$, and $\gamma\eta'$ using a data sample of $27\times10^{6}$ $\psip$ events, and found a large value for the ratio
$R_{\jpsi}=(21.1\pm0.9)\%$ while $R_{\psip}\ll R_{\jpsi}$ with $R_{\psip}<1.8\%$.
The most recent experimental results from the BESIII Collaboration~\cite{Br_bes} confirmed the small value of $R_{\psip}$ and made a first measurement of the
branching fraction $\mathcal{B}(\psip\rightarrow\gamma\po)$ to be $\rm (1.58 \pm 0.4(stat.) \pm 0.13(syst.))\times 10^{-6}$
based on a data sample of $106\times 10^{6}$ $\psip$ events. These results suggest a deviation from the saturation assumption~\cite{SATURATE} and imply that some other mechanisms may be important in $\psip$ radiative decays to a pseudoscalar meson (P).
Reference~\cite{VMD_Mix} discusses decay mechanisms in the framework of the vector-meson dominance model associated with
$\eta_{c}$-$\eta^{(\prime)}$ mixing in order to interpret the difference between $\jpsi$ and $\psip$ radiative decays to a  pseudoscalar meson and predicts
$\mathcal{B}(\psip\rightarrow\gamma\po)=(0.07\sim 0.12)\times10^{-6}$.
Reference~\cite{prePi0} predicts $\mathcal{B}(\psip\rightarrow\gamma\po)\approx 2.19\times10^{-7}$ in the framework of the vector-meson dominance model.

The BESIII detector~\cite{BESIII} has accumulated $(106.9\pm7.5)\times 10^{6}$ and $(341.1\pm2.1)\times10^6$ decays in 2009 and 2012, respectively,
adding up to a total of $448\times10^{6}$ $\psip$ events, corresponding to an integrated luminosity of $509.4$ pb$^{-1}$. The number of $\psip$ decays was determined by counting inclusive hadronic
events~\cite{Npsip09,Npsip12}. The results reported in this paper are based on the complete $\psip$ data sample collected with BESIII and thereby supersede the
previous measurements~\cite{Br_bes}.

\section{\boldmath{BESIII Detector and Monte Carlo Simulation}}

The BESIII detector is described in detail in Ref.~\cite{BESIII}. The detector is cylindrically symmetric and covers 93\% of the solid angle
around the interaction point (IP). The detector consists of four main components: (a) a 43-layer main drift chamber (MDC) provides momentum
measurements for charged tracks with a resolution of 0.5\% at 1~GeV/$c$ in a 1~T magnetic field. (b) a time-of-flight system (TOF) composed of
plastic scintillators has a time resolution of 80 ps (110 ps) in the barrel (endcaps). (c) a 6240-cell CsI(Tl)-crystal electromagnetic
calorimeter (EMC) provides an energy resolution for photons at 1~GeV of 2.5\% (5\%) in the barrel (endcaps).
(d) a muon counter consisting of 9 (8) layers of resistive plate chambers in the barrel (endcaps) within the return yoke of the magnet
provides a position resolution of 2~cm. The electron and positron beams collide with an angle of 22~mrad at the IP in order to separate
the $e^{+}$ and $e^{-}$ beams after the collision.

Monte Carlo (MC) simulations are used to study backgrounds and to determine the detection efficiencies.
The {\sc GEANT4}-based~\cite{GEANT4} simulation software, BESIII Object Oriented Simulation Tool ({\sc BOOST})~\cite{BOOST}, contains a description of the detector geometry and material as well as records of the detector running conditions and performance.
An 'inclusive' MC sample consists of $506\times10^{6}$ generic $\psip$ events, where the $\psip$ is produced by the {\sc kkmc}~\cite{kkmc} generator and its measured decay modes are simulated by {\sc besevtgen}~\cite{evtgen} by setting the branching fractions of known decays according to the Particle Data Group (PDG)~\cite{PDG}, while the remaining unknown decay modes are simulated by {\sc lundcharm}\cite{lundcharm}.
The signal events $\psip\rightarrow\gamma P$ are generated according to the helicity amplitude model HELAMP with the options $(1, 0, -1 ,0)$~\cite{evtgen},
where the options indicate the amplitudes for different partial waves.
In the analysis of $\psip\to\gamma\eta$, the prominent decay mode $\eta\to\gamma\gamma$ is not selected,
since it suffers from the huge Quantum Electrodynamics (QED) background $\EE\to\gamma\gamma$ and, as a consequence, from poor statistical significance.
The other two prominent decay modes $\eta\to\pp\po$ and $\po\po\po$ are selected.
In the analysis of $\psip\to\gamma\eta'$, the $\eta'$ is reconstructed in its decay modes $\eta'\to\pp\eta$ and $\po\po\eta$ with $\eta\to\gamma\gamma$, which have identical final states as those in the analysis of $\psip\to\gamma\eta$.
Many of the systematic uncertainties on the detection efficiency are correlated in the two analyses and will cancel in the measurement of $R_{\psip}$.
In the MC simulation, the decays of $\eta'\to\pi\pi\eta$ and $\eta\to\pi\pi\pi$ are generated according to the measured Dalitz plot distributions~\cite{etapDalitz,
etaDalitz}.
In the analysis of $\psip\to\gamma\po$, the $\po$ signal is reconstructed with its dominant decay mode $\po\to\gamma\gamma$, and the corresponding decay is
described in the MC simulation with a uniform distribution in phase space.

\section{\boldmath{Analysis of $\psip\rightarrow\gamma\eta'/\eta/\pi^0$}}

Charged tracks are reconstructed from hits in the MDC. The polar angle of each track must satisfy $\lvert \cos \theta \rvert<0.93$.
Tracks are required to originate from the IP within $\pm 10$~cm along the beam direction and within 1~cm in the plane perpendicular
to the beam. All selected charged tracks are assumed to be pions.

Photon candidates are reconstructed from isolated clusters in the EMC.
The deposited energy is required to be larger than 25~MeV in the barrel region ($\lvert \cos \theta\rvert<0.80$) or 50~MeV in the endcap
regions $(0.86<\lvert\cos \theta\rvert<0.92)$.
The energy deposited in the nearby TOF counter is included to improve the reconstruction efficiency and energy resolution.
To eliminate clusters associated with charged tracks, the angle extended from the IP between the extrapolated impact point of any charged track in the EMC and a
photon candidate has to be larger than 10 degrees.
For the decays including charged particles in the final states, the timing of EMC clusters with respect to the event start time is used to suppress
electronics noise and photon candidates unrelated to the event.
For the decay with only neutral particles in the final states, the timing requirements are not applied because of the poorly defined start time.

Candidate $\po$ and $\eta$ mesons that do not originate from the $\psip$ radiative decay are reconstructed from pairs of photons.
The invariant mass $M(\gamma\gamma)$ is required to be within $[0.120, 0.150]$  and $[0.522, 0.572]$ GeV/$c^{2}$ for these $\po$ and $\eta$ candidates, respectively.

\subsection{Decay $\psip\rightarrow\gamma\eta'$}

Candidate $\eta'$ mesons are reconstructed in theire decays to $\pp\eta$ and $\po\po\eta$.
We require that there are no additional charged tracks and the number of photon candidates is less than 9.
The photon with the largest energy is regarded as the radiative photon.
Events in the range $0.80< M(\pi\pi\eta) < 1.10$ GeV/$c^2$ are kept for further analysis, and the $\eta'$ signal region is defined as $0.92 < M(\pi\pi\eta) <
0.98$ GeV/$c^2$.
To reduce the backgrounds and to improve the mass resolution, a four-constraint~(4C) kinematic fit is applied to the final state particle candidates, constraining the total four-momentum to the initial values of the colliding beams. The $\chi^{2}_{4C}$ is required to be less than 80.
If more than one possible combination is found in an event, the one with the smallest $\chi^{2}_{4C}$ is retained.
For $\eta^{\prime}\rightarrow\po\po\eta$, we define a variable
$\chi^{2}_{M_{\gamma\gamma}}=(M(\gamma_1\gamma_2)-M_{\po})^{2}/\sigma^{2}_{\po}+(M(\gamma_3\gamma_4)-M_{\po})^{2}/\sigma^{2}_{\po}+(M(\gamma_5\gamma_6)-M_{\eta})^2/\sigma^{2}_{\eta}$,
where $M(\gamma_i\gamma_j)$ is the invariant masses of the photon pair $\gamma_i\gamma_j$, $M_{\po}$ and $M_{\eta}$ are the nominal mass of the $\po$ and $\eta$ taken from the PDG~\cite{PDG}, and $\sigma_{\po}=4.8$~MeV/$c^2$ and $\sigma_{\eta}=8.7$ MeV/$c^2$ are the corresponding mass resolutions.
If there are multiple photon combinations in $\po$ and $\eta$, the one with the least $\chi^{2}_{M_{\gamma\gamma}}$ is retained.

To check the contribution from the continuum process $\EE\rightarrow\gamma\eta'$, we use 44~$\rm pb^{-1}$ of data collected at
a center-of-mass energy $\sqrt{s}=3.65 \rm GeV$~\cite{Npsip09}.
No event within the $\eta'$ signal region passes the $\eta'\rightarrow\po\po\eta$ and $\eta'\rightarrow\pp\eta$ selection criteria.
Therefore the background from non-resonant production is negligible.
For the charged decay mode $\eta'\rightarrow\pp\eta$,
we use the events in the $\eta$ sideband region, $[M_{\eta}-9\sigma, M_{\eta}-6\sigma]$ and $[M_{\eta}+6\sigma, M_{\eta}+9\sigma]$, to check the contribution of non-$\eta$ backgrounds.
The investigation shows that this kind of background distributes uniformly in the region of interest in the $\pp\eta$ invariant mass spectrum.
A study of the inclusive $\psip$ MC sample reveals that the channels $\psip\rightarrow\pp\jpsi$ with $\jpsi\rightarrow\gamma\eta$ and
$\psip\rightarrow\gamma\eta\pp$ are the dominant backgrounds with an $\eta$ in the final state.
The channels $\psip\rightarrow\pi^{0}\pi^{0}\jpsi$ with $\jpsi\rightarrow\gamma\eta$ and $\psip\rightarrow\eta\jpsi$ with $\jpsi\rightarrow\gamma\eta$ and
$\eta\rightarrow\po\po\po$ are the main backgrounds in the neutral mode, $\eta'\rightarrow\po\po\eta$.
The contribution from $\psip\rightarrow\gamma\eta\po\po$ is negligible because of the small branching fraction.
All of the above backgrounds distribute smoothly and do not produce peaks in the vicinity of the $\eta'$ signal in the $\pi\pi\eta$ invariant mass spectrum,
$M(\pi\pi\eta)$.

\begin{figure*}[htbp]
\begin{flushleft}
    \includegraphics[width=0.47\textwidth]{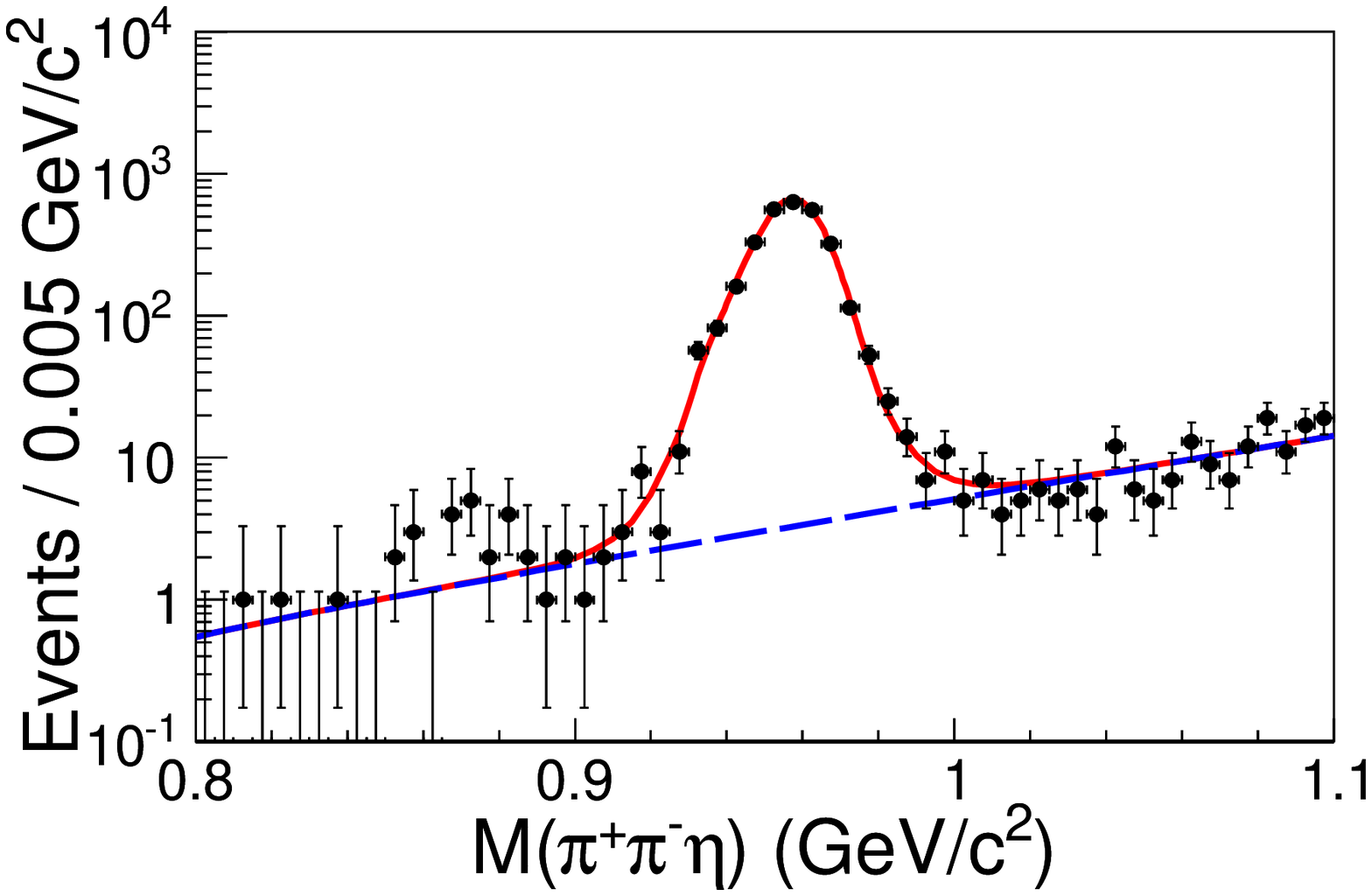}
    \includegraphics[width=0.47\textwidth]{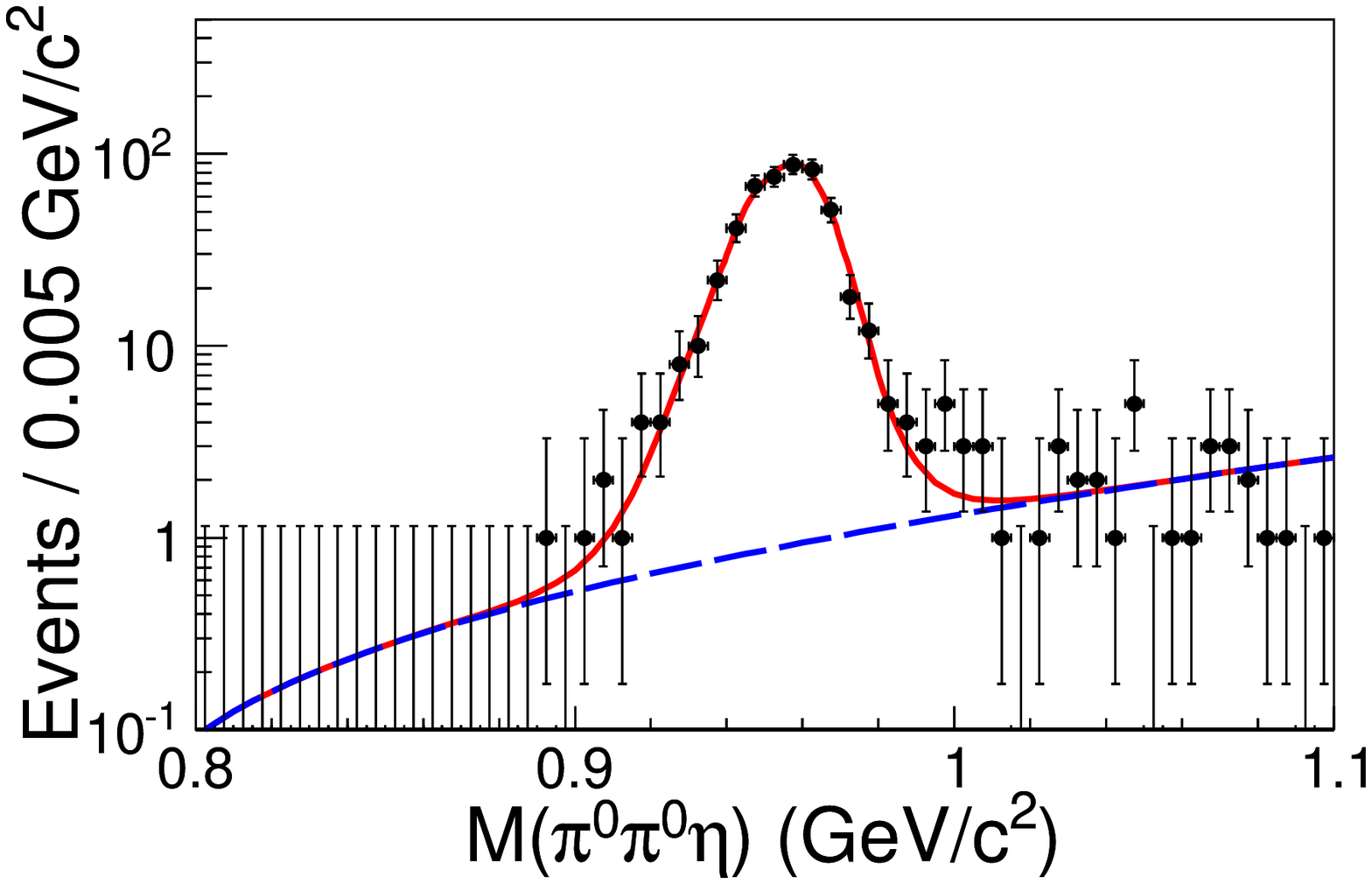}
\end{flushleft}
\caption{(color online) Projection of the simultaneous fit to the invariant mass of $M(\pi^{+}\pi^{-}\eta)$ (left) and $M(\pi^{0}\pi^{0}\eta)$ (right).
Dots with error bars show data, the red solid curves show the total fit result, and the blue dashed lines represent the background contributions. }\label{fig:FitResult_etap}
\end{figure*}

Figure~\ref{fig:FitResult_etap} shows the $M(\pi\pi\eta)$ distributions for selected $\pp\eta$~(left) and $\po\po\eta$~(right) candidates.
Prominent $\eta'$ signals are observed  in both decay modes.
To determine the signal yield, a simultaneous unbinned maximum likelihood fit is performed to the mass spectra of both decay modes.
The ratio of the number of $\pp\eta$ signal events to that of $\po\po\eta$ signal events is fixed to be
$\frac{\epsilon_{\pp\eta} \cdot \mathcal{B}(\eta'\rightarrow\pp\eta)}{\epsilon_{\po\po\eta} \cdot \mathcal{B}(\eta'\rightarrow\po\po\eta) \cdot
\mathcal{B}^{2}(\po\rightarrow\gamma\gamma)}$,
where $\mathcal{B}(\eta'\rightarrow\pi^{+/0}\pi^{-/0}\eta)$ and $\mathcal{B}(\po\rightarrow\gamma\gamma)$ are the branching fractions
taken from the PDG~\cite{PDG}, and $\epsilon_{\pp\eta}=30.8\%$ and $\epsilon_{\po\po\eta}=9.0\%$ are the respective
reconstruction efficiencies determined from signal MC simulations.
In the fit, the signals are described by the MC-determined shapes convolved with a Gaussian function representing the remaining discrepancy
between the data and MC simulation, where the parameters of the Gaussian function are left free in the fit.
The backgrounds are described with an ARGUS~\cite{Argus} function with the threshold parameter fixed slightly below the kinematical limit to take into account the finite experimental resolution on the $\eta$ and $\po$ masses.
The fit results are shown in Fig.~\ref{fig:FitResult_etap}, and the goodness-of-fit is $\chi^{2}/\rm d.o.f=74.7/48$.
The signal yield of $\psip\rightarrow\gamma\eta'$ corrected for reconstruction efficiency and the subsequent decay branching fractions is $56053.5\pm980.8$, where the error is statistical only.

\subsection{Decay $\psip\rightarrow\gamma\eta$}

The $\eta$ candidates in the decay $\psip\rightarrow\gamma\eta$ are reconstructed using the prominent decay modes $\pp\po$ and $\po\po\po$.
The event selection is similar to that of the $\psip\rightarrow\gamma\eta'$ analysis, since they have the same final states,
except that we do not apply the requirement of the angle between charged tracks and isolated photons  because of the higher momentum of the $\eta$ candidates. As a
consequence, the photons from the $\po$ decay can be close to the charged pions.

For $\psip\rightarrow\gamma\eta$ with $\eta\rightarrow\pp\po$, the main backgrounds come from the tail of continuum process $\EE\rightarrow\gamma_{ISR}\omega$, which is studied using the data taken at $\sqrt{s}=3.65$ GeV.
The $\pp\po$ mass spectrum for the continuum process is flat and the expected number of events is $251.6\pm58.8$.
The backgrounds from $\psip$ decays are examined with the 506 M inclusive MC sample.
Only one such event survives, and this class of background is therefore ignored.
For $\psip\rightarrow\gamma\eta$ with $\eta\rightarrow\po\po\po$, the possible peaking background is from the decays $\psip\rightarrow\gamma\chi_{cJ}$ with $\chi_{cJ}\rightarrow\eta(\pi^{0}\pi^{0}\pi^{0})\eta(\gamma\gamma)$, which is expected to produce $0.6\pm0.1$ events in the signal region according to the MC simulation.
Therefore, this source of background can be ignored.
The background from the continuum process, studied with the data taken at $\sqrt{s}=3.65$ GeV, is expected to contribute less than one event, and is also ignored.

\begin{figure*}[htbp]
\begin{flushleft}
    \includegraphics[width=0.47\textwidth]{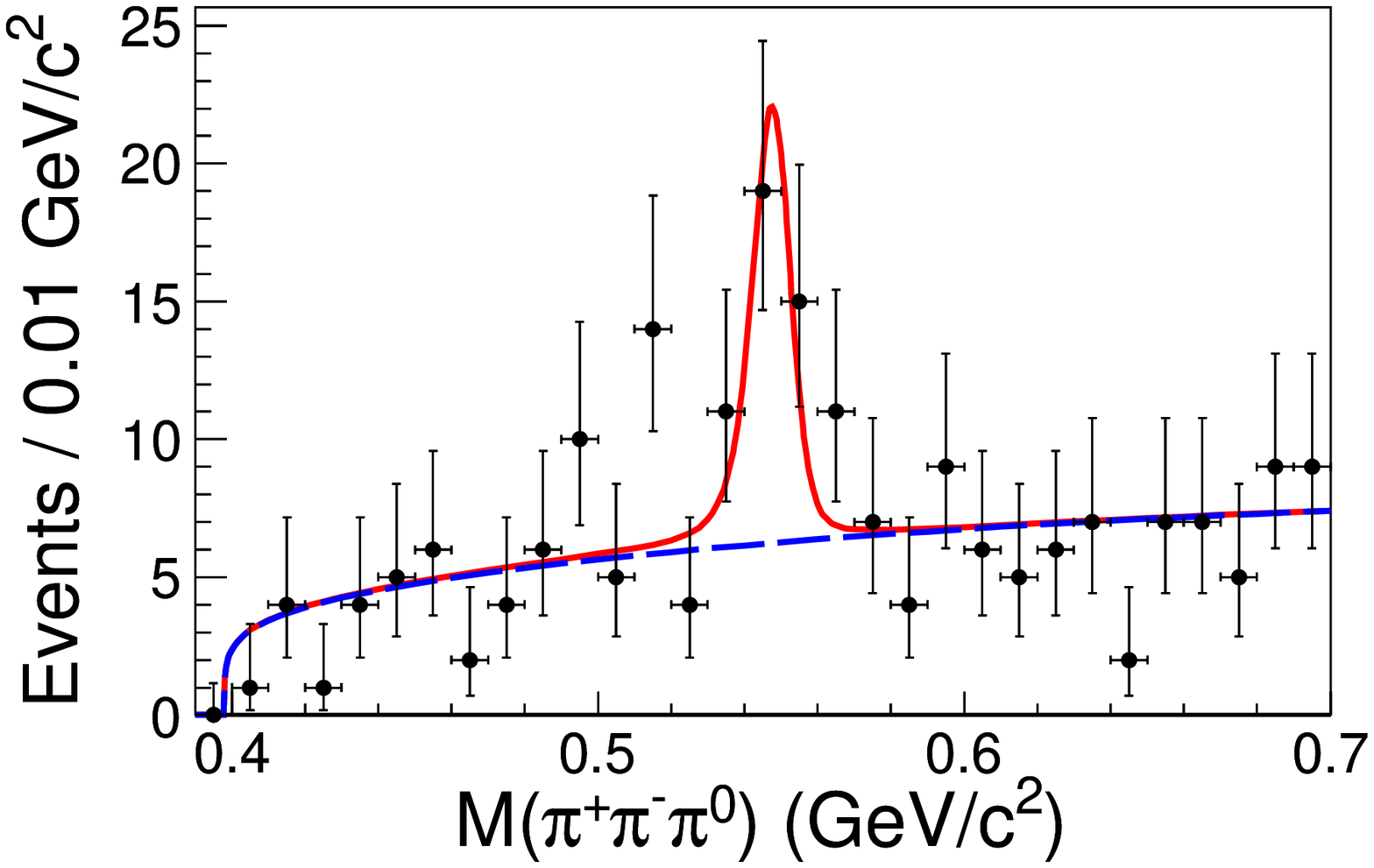}
    \includegraphics[width=0.47\textwidth]{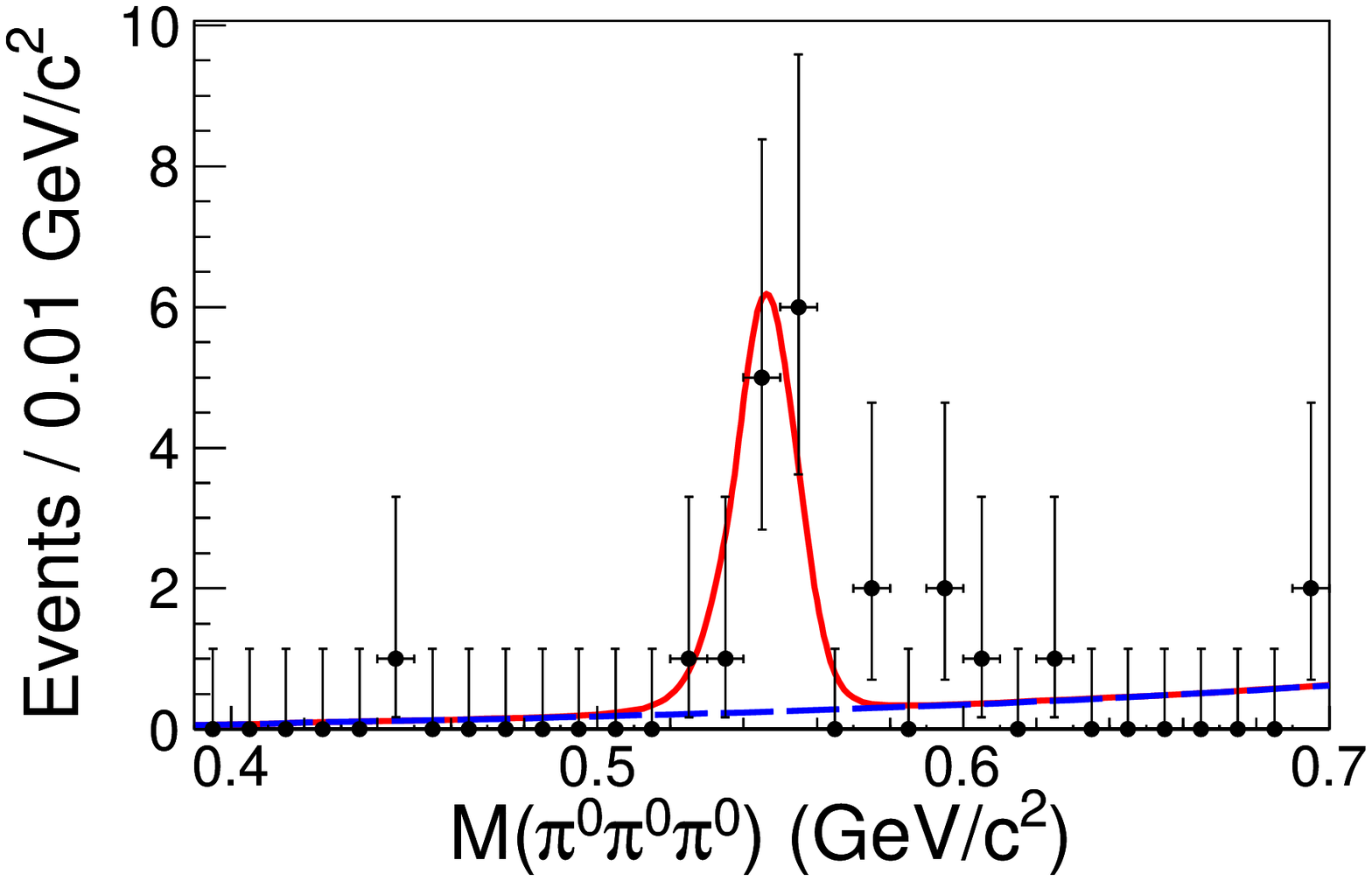}
\end{flushleft}
\caption{(color online) Projection of the simultaneous fit to the invariant mass of $M(\pi^{+}\pi^{-}\po)$ (left) and $M(\pi^{0}\pi^{0}\po)$ (right).
Dots with error bars represent the data, the red solid curves show the total fit result, and the blue dashed lines correspond to the background contributions.
}\label{fig:FitResult_eta}
\end{figure*}

The $M(\pi\pi\pi)$ invariant mass is used to determine the signal yield of $\psip\rightarrow\gamma\eta$.
Figure~\ref{fig:FitResult_eta} shows the distribution of $M(\pp\po)$  (left)
and $M(\po\po\po)$ (right) for selected $\pp\po$ and $\po\po\po$ candidates, respectively.
A clear peak for the $\eta$ signal is seen in both $M(\pi\pi\pi)$ distributions.
The signal region is defined as [0.522, 0.572]~GeV/$c^{2}$ and the fit range is
[0.380, 0.700]~GeV/$c^{2}$.
A simultaneous unbinned maximum likelihood fit is applied to the $M(\pp\po)$ and $M(\po\po\po)$ spectra. The ratio of the number of $\pp\po$ signal events to that of $\po\po\po$ signal events
is fixed at $\frac{\epsilon_{\pp\po} \cdot \mathcal{B}(\eta\rightarrow\pp\po)}{\epsilon_{\po\po\po} \cdot \mathcal{B}(\eta\rightarrow\po\po\po) \cdot
\mathcal{B}^{2}(\po\rightarrow\gamma\gamma)}$,
where $\mathcal{B}(\eta\rightarrow\pi\pi\pi)$ and $\mathcal{B}(\po\rightarrow\gamma\gamma)$ are the branching fractions quoted from the PDG~\cite{PDG};
$\epsilon_{\pp\po}=29.0$\% and $\epsilon_{\po\po\po}=12.1$\% are the reconstruction efficiencies determined from the signal MC simulations.
In the fit, the signal is described with the MC-determined shape convolved with a Gaussian function, where the parameters of the Gaussian function are fixed to
those obtained in the simultaneous fit for $\psip\rightarrow\gamma\eta'$, which has the same final state and higher statistics.
The background is described with an ARGUS function, where the threshold parameter is fixed to the sum of the mass of the three pion.
The fit results are shown in Fig.~\ref{fig:FitResult_eta} as solid curves.
The signal yield of $\psip\rightarrow\gamma\eta$ after correcting for efficiency and the subsequent decay branching fractions is $382.5\pm78.9$, where the uncertainty is statistical only.
The goodness-of-fit is $\chi^{2}/\rm d.o.f=16.6/11$, using only bins with at least seven events.
The statistical significance of $\psip\to\gamma\eta$ is 7.3$\sigma$ by comparing the likelihood values of the fits with or without the $\eta$ signal included
($\Delta(\rm ln \mathcal{L})=27.0$) and taking into account the change in the number of degrees of freedom ($\Delta(\rm d.o.f)=1$).

\subsection{Decay $\psip\rightarrow\gamma\pi^0$}

To select candidate events for the decay $\psip\rightarrow\gamma\pi^0$, the events are required to have exactly three reconstructed showers and no good charged tracks.
To suppress the QED background $\EE\rightarrow\gamma\gamma(\gamma_{ISR})$, only photons in the barrel region ($|\cos\theta|<0.8$)
are accepted.
A 4C kinematic fit is performed, and the $\chi^{2}_{4C}$ is required to be less than 40.
The most energetic photon is regarded as the radiative one.
To further suppress the QED background, the cosine of the helicity angle of the $\po$, which is defined as the angle between the momentum direction of the more
energetic photon in the $\pi^{0}$ rest frame and the $\pi^{0}$ momentum in the $\psip$ rest frame, is required to be
less than 0.7.

Based on a study of the continuum data at $\sqrt{s}=3.65$ GeV and the inclusive MC sample, we find that the $\EE\rightarrow\gamma\gamma(\gamma_{ISR})$ processes
contaminate the signal.
One of the photons converts into an $\EE$ pair, which are misidentified as two photons if the track finding algorithm fails.
To remove this kind of background, we require fewer than eight hits in the MDC in the region between the two radial lines connecting the
IP and the two shower positions in the EMC.
According to MC studies, almost all of peaking background caused by the gamma conversion process in the $\EE\rightarrow\gamma\gamma$ events can be rejected with only a 2.7\% loss in the signal efficiency.
The other backgrounds are the decays $\psip\rightarrow\gamma\chi_{cJ}$ ($J=0,2$), with $\chi_{cJ}\rightarrow\po\po$, which produce a peak in the signal region in the two-photon invariant mass.
According to MC simulations and using the well-measured branching fractions quoted in the PDG~\cite{PDG},
the background is expected to be $36.7\pm1.7$ events.

\begin{figure}[htbp]
\begin{center}
    \includegraphics[width=0.5\textwidth]{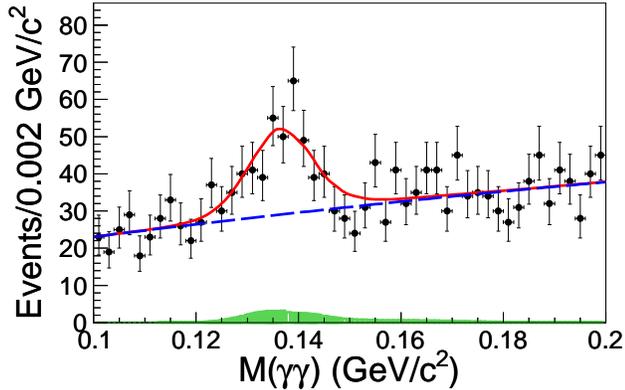}
\end{center}
\caption{(color online) Unbinned maximum likelihood fit to the $M(\gamma\gamma)$ spectrum for the decay $\psip\rightarrow\gamma\po$.
Dots with error bars show data.
The red solid curve shows the result of the fit, the blue dashed line represents the contribution of
the QED background, and the green shaded histogram depicts the peaking background from $\chi_{cJ}\to\pi^0\pi^0$ decays. }\label{fig:FitResult_pi0}
\end{figure}

Figure~\ref{fig:FitResult_pi0} shows the $M(\gamma\gamma)$ spectrum for selected $\psip\rightarrow\gamma\po$ candidates.
A clear peak from the $\po$ signal is observed.
An unbinned maximum likelihood fit to the $M(\gamma\gamma)$ distribution is performed to detemine the signal yield.
The fit function consists of three components representing the signal, a smooth background from $\EE\rightarrow\gamma\gamma(\gamma_{ISR})$ events,
and a contribution from $\psip\rightarrow\gamma\chi_{cJ}$ decays with $\chi_{cJ}\rightarrow\po\po$.
The signal is modeled by a MC simulated shape convoluted with a Gaussian function representing the resolution difference between the
MC simulation and the data. The parameters of the Gaussian function are left free in the fit.
The shape parameters of the smooth background are determined from the MC simulation and the magnitude is determined by the fit to data.
The size and shape of the contribution from $\psip\rightarrow\gamma\chi_{cJ}$ decays with $\chi_{cJ}\rightarrow\po\po$ are fixed according to the expectation from
MC studies.
The results of the maximum likelihood fit are shown in Fig.~\ref{fig:FitResult_pi0} and the goodness-of-fit
is $\chi^{2}/\rm d.o.f=40.6/46$.
The signal yield after correcting for the efficiency, which is 36.8\% according to the MC simulation, and the subsequent decay branching fraction is $423.4\pm71.4$, and the statistical significance of the $\po$
signal is $6.7\sigma$ ($\Delta(\rm ln \mathcal{L})=26.1,\Delta(\rm d.o.f)=3$).

In the above three analyses, the branching fractions are obtained using the signal yields $N^{\rm cor}_{\rm sig}$, corrected for the detection efficiency and the subsequent
branching fraction, and the total number
of $\psip$ events $N^{\psip}_{\rm tot}$ according to $\mathcal{B}=\frac{N^{\rm cor}_{\rm sig}}{N_{\rm tot}^{\psip}}$.
The results are summarized in Table~\ref{tab:result}.

\begin{table*}[htbp]
\caption{A comparison of our results with previously published BESIII measurements. $N^{\rm cor}_{\rm sig}$ is the signal yield, corrected for efficiency and subsequent branching fractions, as obtained from the fits. The statistical significances are presented as well.} \label{tab:result}
\begin{tabular}{c | c | c | c | c  }
  \hline
  \hline
  Decay mode & Significance  &  $N^{\rm cor}_{\rm sig}$ & $\mathcal{B}(\psip\rightarrow\gamma\eta'/\eta/\po)$ & Previous results from BESIII~\cite{Br_bes}   \\
  \hline
  $\psip\rightarrow\gamma\eta'$   &   $>10\sigma$   &  $56053.5\pm980.8$  & $(125.1\pm2.2\pm6.2)\times10^{-6}$   & $(126 \pm 3 \pm 8)\times10^{-6}$             \\
  $\psip\rightarrow\gamma\eta$    & $7.3\sigma$ &  $382.5\pm78.9$ & $(0.85\pm0.18\pm0.04)\times10^{-6}$  & $(1.38 \pm 0.48 \pm 0.09)\times10^{-6}$       \\
  $\psip\rightarrow\gamma\pi^{0}$ & $6.7\sigma$ & $423.4\pm71.4$ & $(0.95\pm0.16\pm0.05)\times10^{-6}$  & $(1.58 \pm 0.40 \pm 0.13)\times10^{-6}$       \\
  \hline

  \hline
\end{tabular}
\end{table*}

\section{\boldmath Systematic uncertainties}\label{Sys_err}

The main sources of systematic uncertainty in the branching fraction measurements stem from the data-simulation differences in the track reconstruction efficiency,
the photon detection efficiency, the $\eta$ and $\po$ reconstruction efficiency, and the kinematic fit, and the uncertainties from the related branching fraction in the cascade decays, the number of hits in the MDC and the number of photons required in $\psip\rightarrow\gamma\po$,
 the fit procedure, and the decay models of the $\eta'$ and $\eta$ in the MC simulation, as well as the total number of $\psip$ events.

The uncertainty due to the charged track reconstruction is studied with the control sample
$\psip\rightarrow\pp\jpsi,\jpsi\rightarrow\LL$, and is 1\% per track~\cite{track}.
The uncertainty for the photon detection efficiency is 1\% for each photon on average, obtained by studying the
control sample $\jpsi\rightarrow\rho^{0}\po$ ~\cite{Npsip13}.
In studying the $\psip\rightarrow\gamma\po$ mode, only the photons within the barrel EMC region are used, which significantly improves the systematic uncertainty,
estimated to be 0.5\% per photon.
For the reconstruction of the $\eta$ and $\po$ mesons from their two-photon decay mode,
the systematic uncertainty is 1.0\% per meson~\cite{track}.

The uncertainty associated with the kinematic fit arises from the inconsistency of the track helix parameter and photon between the data and the MC simulation.
For the decay processes including charged tracks in the final state, we correct the three helix parameters ($\phi_{0}, \kappa$ and $\rm tan\lambda$) of the charged
tracks in the signal MC samples to reduce this deviation,
where the correction factors are obtained by comparing their pull distributions in a control sample of $\psip\rightarrow K^{+} K^{-} \pp$ between data and MC
simulation, as described in Ref.~\cite{CRfactor}.
We also estimate the detection efficiency without the helix parameter corrections, and the resulting change in the detection efficiency, 1\%, is taken as the
systematic uncertainty.
For the decay processes with purely neutral particles in the final states, the uncertainty associated with the kinematic fit is studied using the decay
$\EE\rightarrow\gamma\gamma\gamma_{ISR}$ as the control sample. The ratios of the number of events with and without the kinematic fit are obtained.
The difference in the ratios between the data and MC simulations, 2.0\%, is considered as the systematic uncertainty due to the kinematic fit.

In the analysis of $\psip\rightarrow\gamma\po$, the additional requirement on the number of hits in the MDC is applied to suppress the dominant background
$\EE\rightarrow\gamma\gamma(\gamma_{ISR})$.
The corresponding efficiency is studied with the control sample $\psip\rightarrow\gamma\chi_{c2}$ with $\chi_{c2}\rightarrow\gamma\gamma$, which has same final
state as the signal process of interest.
The plane region used to count the MDC hits in the control sample is larger than that in the $\psip\rightarrow\gamma\po$ decay due to the smaller Lorentz boost of
the $\gamma\gamma$ system, and as a consequence more MDC hits from noise will be counted in the control sample.
To minimize this effect, we normalize the MDC hits according to area by assuming the noise is distributed uniformly over the MDC.
The difference in the efficiency between the data and MC simulation is 1\%, which is taken as a systematic uncertainty.
Analogously, the selection efficiency for the photon multiplicity requirement, $N_{\gamma}=3$, is studied with the same control samples. The resulting difference
between the data and MC simulation, 3.1\%, is regarded as the systematic uncertainty.

The sources of systematic uncertainty in the fit procedures include the fit range and the background.
The uncertainty associated with the fit range is estimated by varying the fit range by $\pm 0.01 \rm GeV/c^{2}$; the largest resulting change in the signal
yields with respect to the nominal values are taken as the uncertainties.
In the analysis of $\psip\rightarrow\gamma\eta'$ and $\gamma\eta$, the uncertainties related to the background shape are estimated by replacing the ARGUS functions
with polynomials functions in the fit.
The resulting changes in the signal yields with respect to the nominal values are considered as the systematic uncertainties.
In the analysis of $\psip\rightarrow\gamma\po$, the peaking backgrounds from the $\psip\rightarrow\gamma\chi_{c0,2}$ decay are included in the fit and the
corresponding strengths  are fixed to the values estimated from the MC simulation, incorporating the branching fraction from the PDG~\cite{PDG}.
To evaluate the systematic uncertainty, we change the strength of the peaking background by $\pm 1$ times the standard deviations of the background strength, and
repeat the fits.
The larger change of the signal yield, 2.1\%, is taken as the systematic uncertainty.

In the MC simulation, we generate the $\eta'\to\pp\eta$ and $\eta\to\pp\po$ signals according to Ref.~\cite{etapDalitz,etaDalitz}.
We vary the parameters within $\pm1$ times the standard deviation in the generator. The changes in the reconstruction efficiency, 0.6\% for the $\eta'$ mode and
0.4\% for the $\eta$ mode, are taken as the systematic uncertainties.

The uncertainty in the total number of $\psip$ decays is estimated to be $0.6\%$~\cite{Npsip09,Npsip12}.
The uncertainties related to the branching fractions in the cascade decays are quoted from the PDG~\cite{PDG}.

Table~\ref{tab:Sys_Err} summarizes the various systematic uncertainties for the decays of interest.
The overall systematic uncertainties are obtained by adding the individual uncertainties in quadrature, taking into account the correlation between the different
decay modes.
Compared to the previous BESIII measurements~\cite{Br_bes}, improved systematical uncertainties are obtained due to the improved measurement of the total number
of $\psip$ events and better fits to the corresponding invariant mass to determine the signal yields.

\begin{table*}[ht]
\caption{A summary of all sources of systematic uncertainties (in \%) in the branching fraction measurements.
The ellipsis $''...''$ indicates that the uncertainty is negligible or not applicable.} \label{tab:Sys_Err}
\begin{tabular}{c | c  c | c  c | c }
  \hline
  \hline
   & $\gamma\eta'(\pi^{+}\pi^{-}\eta)$ & $\gamma\eta'(\pi^{0}\pi^{0}\eta)$ & $\gamma\eta(\pi^{+}\pi^{-}\pi^{0})$ & $\gamma\eta(\pi^{0}\pi^{0}\pi^{0})$ &
   $\gamma\pi^{0}(\gamma\gamma)$ \\
  \hline
  Tracking                                    & 2.0 & ... & 2.0 & ... & ... \\
  Photon detection                            & 3.0 & 7.0 & 3.0 & 7.0 & 1.5 \\
  $\eta$ reconstruction                       & 1.0 & 1.0 & ... & ... & ... \\
  $\po$ reconstruction                        & ... & 2.0 & 1.0 & 3.0 & ... \\
  Kinematic fit                               & 1.0 & 2.0 & 1.0 & 2.0 & 2.0 \\
  Branching fraction                          & 1.7 & 3.7 & 1.2 & 0.7 & 0.0 \\
  \hline
  Requirement on MDC hits             &\multicolumn{2}{c|}{ ... } &\multicolumn{2}{c|}{ ... } & 1.0 \\
  Number of good photons              &\multicolumn{2}{c|}{ ... } &\multicolumn{2}{c|}{ ... } & 3.1 \\
  Fitting range                       &\multicolumn{2}{c|}{ 0.2 } &\multicolumn{2}{c|}{ 1.2 } & 3.5 \\
  Background shape                    &\multicolumn{2}{c|}{ 0.3 } &\multicolumn{2}{c|}{ 1.2 } & ... \\
  Background estimation               &\multicolumn{2}{c|}{ ... } &\multicolumn{2}{c|}{ ... } & 2.1 \\

  $\eta/\eta'$ decay models           &\multicolumn{2}{c|}{ 0.6 } &\multicolumn{2}{c|}{ 0.4 } & ... \\
  $\psip$ total number                &\multicolumn{2}{c|}{ 0.6 } &\multicolumn{2}{c|}{ 0.6 } & 0.6 \\
  \hline
  total                               &\multicolumn{2}{c|}{ 4.9 } &\multicolumn{2}{c|}{ 5.3 } & 5.6 \\
  \hline
\end{tabular}
\end{table*}

\section{\boldmath Summary}
By analyzing the data sample of $448\times10^{6}$ $\psip$ events collected at $\sqrt{s}=3.686$ GeV with the BESIII detector,
we observe clear signals of $\psip$ decays to $\gamma\eta',\gamma\eta$, and $\gamma\pi^{0}$.
The statistical significance of $\psip\rightarrow\gamma\eta$ and $\gamma\po$ are $7.3\sigma$ and $6.7\sigma$, respectively, and the decay branching fractions are
measured with much improved precision, superseding the previous BESIII measurement~\cite{Br_bes}.
A comparison of these results to those in Ref.~\cite{Br_bes} is shown in Table~\ref{tab:result}.
The branching fraction of $\psip\rightarrow\gamma\eta'$ is consistent with the previous measurement but with improved precision,
while those of $\psip\rightarrow\gamma\eta$ and $\gamma\po$ are lower than the previous results, but are consistent within $1\sigma$.

The ratio of branching fractions for $\psip$ radiative decays to $\eta$ and $\eta'$ is calculated to be
$R_{\psip}=(0.66\pm0.13\pm0.02)\%$.
This is about 30 times smaller than the corresponding ratio from $\jpsi$ radiative decays,
$R_{\jpsi}=(21.4\pm0.9)\%$.
The large difference in the ratios of branching fractions between $\jpsi$ and $\psip$ radiative decays can be explained by the approach proposed in
Ref.~\cite{VMD_Mix}.
However, the predicted branching fraction of $\psip\rightarrow\gamma\po$ in Ref.~\cite{VMD_Mix}, $\mathcal{B}(\psip\rightarrow\gamma\po)=(0.66\sim1.15)\times10^{-7}$,
turns out to be one order smaller than that in this measurement.
Further investigations are necessary to understand the discrepancy.
The results presented in this paper provide an ideal benchmark for testing various theoretical models of radiative decays
of $c\bar{c}$ bound states.

The BESIII collaboration thanks the staff of BEPCII and the IHEP computing center for their strong support. This work is supported in part by National Key Basic Research Program of China under Contract No. 2015CB856700; National Natural Science Foundation of China (NSFC) under Contracts Nos. 11235011, 11335008, 11425524, 11625523, 11635010; the Chinese Academy of Sciences (CAS) Large-Scale Scientific Facility Program; the CAS Center for Excellence in Particle Physics (CCEPP); Joint Large-Scale Scientific Facility Funds of the NSFC and CAS under Contracts Nos. U1332201, U1532257, U1532258; CAS under Contracts Nos. KJCX2-YW-N29, KJCX2-YW-N45, QYZDJ-SSW-SLH003; 100 Talents Program of CAS; National 1000 Talents Program of China; INPAC and Shanghai Key Laboratory for Particle Physics and Cosmology; German Research Foundation DFG under Contracts Nos. Collaborative Research Center CRC 1044, FOR 2359; Istituto Nazionale di Fisica Nucleare, Italy; Joint Large-Scale Scientific Facility Funds of the NSFC and CAS; Koninklijke Nederlandse Akademie van Wetenschappen (KNAW) under Contract No. 530-4CDP03; Ministry of Development of Turkey under Contract No. DPT2006K-120470; National Natural Science Foundation of China (NSFC) under Contract No. 11505010; National Science and Technology fund; The Swedish Resarch Council; U. S. Department of Energy under Contracts Nos. DE-FG02-05ER41374, DE-SC-0010118, DE-SC-0010504, DE-SC-0012069; University of Groningen (RuG) and the Helmholtzzentrum fuer Schwerionenforschung GmbH (GSI), Darmstadt; WCU Program of National Research Foundation of Korea under Contract No. R32-2008-000-10155-0.

\end{CJK*}

\begin{thebibliography}{**}

\bibitem{Scharre:1980yn}
  D.~L.~Scharre {\it et al.},
  Phys.\ Rev.\ D {\bf 23}, 43 (1981).
\bibitem{mixing1} H.~Fritzsch and J.~D.~Jackson, Phys. Lett. B {\bf66}, 365 (1977).
\bibitem{mixing2} K.~T.~Chao, Nucl. Phys. B {\bf335}, 101 (1990).
\bibitem{VMD1} V.~L.~Chernyak and A. R. Zhitnitsky, Phys. Rep. {\bf112}, 173 (1984).
\bibitem{VMD2} G.~W.~Intemann, Phys. Rev. D {\bf27}, 2755 (1983).
\bibitem{VMD_Mix} Q.~Zhao, Phys. Lett. B {\bf697}, 52 (2011).
\bibitem{R1R2} V.~L.~Chernyak and A.R. Zhitmitsky, Phys. Rep. {\bf112}, 173(1984).

\bibitem{PDG} C.~Patrignani {\it et al.} [Particle Data Group],Chin. Phys. C, 40, 100001 (2016).
\bibitem{SATURATE} T.~Feldmann, P.~Kroll and B.~Stech, Phys.\ Lett.\ B {\bf 449}, 339 (1999).

\bibitem{Br_cleo} T.~K.~Pedlar {\it et al.} [CLEO Collaboration], Phys. Rev. D {\bf79}, 111101 (2009).
\bibitem{Br_bes} M.~Ablikim {\it et al.} [BESIII Collaboration], Phys.\ Rev.\ Lett.\  {\bf 105}, 261801 (2010).
\bibitem{prePi0}  J.~L.~Rosner, Phys.\ Rev.\ D {\bf 79}, 097301(2009).
\bibitem{BESIII} M.~Ablikim {\it et al.} [BESIII Collaboration], Nucl. Instrum. Meth. A {\bf614}, 345 (2010).
\bibitem{Npsip09}  M. Ablikim {\it et al.} [BESIII Collaboration], Chin. Phys. C {\bf37}, 063001 (2013).
\bibitem{Npsip12} With the same method (as described in Ref.~\cite{Npsip09}), the preliminary number of the $\psip$ events taken in 2009 and 2012 is
    determined to be $448\times 10^{-6}$ with an uncertainty of 0.6\%.
\bibitem{GEANT4} S.~Agostinelli {\it et al.} [GEANT4 Collaboration], Nucl. Instrum. Meth. A {\bf506}, 250 (2003).
\bibitem{BOOST} Z. Y. Deng et al., Chin. Phys. C {\bf 30}, 371 (2006).


\bibitem{kkmc} S.~Jadach, B.~F.~L. Ward, and Z.~Was, Comput. Phys. Commun. {\bf130}, 260 (2000); Phys. Rev. D {\bf63}, 113009 (2001).
\bibitem{evtgen} D.~J.~Lange, Nucl. Instrum. Meth. A {\bf462}, 152 (2001); R.~G.~Ping, Chin. Phys. C {\bf32}, 599(2008).
\bibitem{lundcharm} J.~C.~Chen, G. S. Huang, X. R. Qi, D. H. Zhang, and Y.~S.~Zhu, Phys.\ Rev.\ D {\bf 62}, 034003 (2000).
\bibitem{etapDalitz}  M.~Ablikim {\it et al.} [BESIII Collaboration], 'Measurement of the matrix elements for the decays $\eta'\to\eta\pi\pi$', publication to be submitted to Phys.Rev.D.
\bibitem{etaDalitz}  M.~Ablikim {\it et al.} [BESIII Collaboration], Phys.\ Rev.\ D {\bf 92}, 012014 (2015).
\bibitem{Argus} H.~Albrecht {\it et al.} [ARGUS Collaboration], Phys.\ Lett.\ B {\bf 241}, 278 (1990).
\bibitem{track}      M.~Ablikim {\it et al.} [BESIII Collaboration], Phys. Rev. Lett. {\bf105}, 261801 (2010).
\bibitem{Npsip13}  M.~Ablikim {\it et al.} [BESIII Collaboration], Phys. Rev. Lett. {\bf116}, 251802 (2016).
\bibitem{CRfactor}   M.~Ablikim {\it et al.} [BESIII Collaboration], Phys. Rev. D {\bf87}, 012002 (2013).



\end{thebibliography}
\end{document}